# A Comparison Between Long Short-Term Memory and Hidden Markov Model to Predict Productivity of Maize in Nigeria

Nkemnole, E. B.[1]  Adoghe V. O.[2]

[1,2] Department of Statistics University Lagos,
enkemnole@unilag.edu.ng, adoghevictor5@gmail.com

**Abstract**

Due to population increase and import constraints, maize, a key cereal crop in Africa, is experiencing a boom in demand. Given this, the study's focus is on determining how maize output in Nigeria interacts with various climatic factors, particularly rainfall and temperature. The Hidden Markov Model (HMM) and the Long Short-Term Memory neural network (LSTM) are compared in this context to assess their performance. A variety of performance indicators, such as correlation, mean absolute percentage error (MAPE), standard error of the mean (SEM), and mean square error (MSE), are used to evaluate the models. The outcomes show that the HMM performs better than the LSTM, with an RMSE of 1.21 and a MAPE of 12.98 demonstrating greater performance. Based on this result, the HMM is then used to forecast maize yield while taking the effects of temperature and rainfall into account. The estimates highlight the possibility for increasing local output by demonstrating a favorable environment for maize planting in Nigeria. In order to help the Nigerian government in its efforts to increase maize production domestically, these studies offer useful insights.

**Keywords:** Hidden Markov Model, LSTM, Time Series, Maize, Baum-Welch Algorithm

**Introduction**

Maize (Zea mays L.) is a vital cereal crop in Africa and the developing world, including Nigeria, crucial for food security and poverty reduction (Olaniyi and Adewale, 2012). However, challenges such as low productivity and limited adoption of improved technologies hinder maize production in Nigeria (Olaniyi and Adewale, 2012). Behind the so-called maize issue is a larger issue with the provision of staple foods, particularly in urban areas. The amount of locally produced staple food is not enough to satisfy demand given the technological state of rural and urban areas today and the pricing structure of basic food items. In order to satisfy local demand, a number of agricultural items exported before I960 have been imported in recent years.

Maize, one of Nigeria's most popular food crops is consumed by millions of Nigerians and is also used to make animal feed. On July 13[th], 2020, the central bank of Nigeria directed all authorized dealers to stop processing Form M for maize importation with immediate effect. The unpredictable wealth conditions and the negative effect of the covid-19 pandemic have been added to the unknown factors that as caused a shortfall in maize production.

M. Yahaya and A. Lawal (2021) did a stochastic model for rice yield forecast employing the Hidden Markov Model (HMM) on data pertaining to Niger state. However, in most cases, the production is hidden or invisible, and each state randomly generates one out of every k observation that is visible.

However, the present literature did not compare the model used with another model with a dataset emanating from Niger state. Therefore, this study will demonstrate the comparative of the prediction of Hidden Markov models (HMM) and Long Short-Term Memory Neural Networks (LSTM-NN) with yearly datasets of maize production cases general Nigeria, to provide necessary information to the government to boost maize production in Nigeria.

**Hidden Markov Model**

The Hidden Markov Model (HMM), first introduced in 1957, finds extensive applications in various industries such as management, engineering, medicine, signal processing, and production (Mac Donald and Zucchini, 1997; Cappe et al., 2005). HMM is a discrete stochastic process $(X_t, Y_t)_{t \geq 0}$, composed of an underlying unobservable process $(X_t)$ and an observed process $(Y_t)$ generating independent random observations. The components of an HMM include states, transitions, observations, and probabilistic behavior (Fig 2.1). The HMM's parameters $(A, B, \prod)$, representing the state transition probability matrix, state emission probability matrix, and initial state probability matrix, fully describe its behavior (Mac Donald and Zucchini, 1997; Cappe et al., 2005).

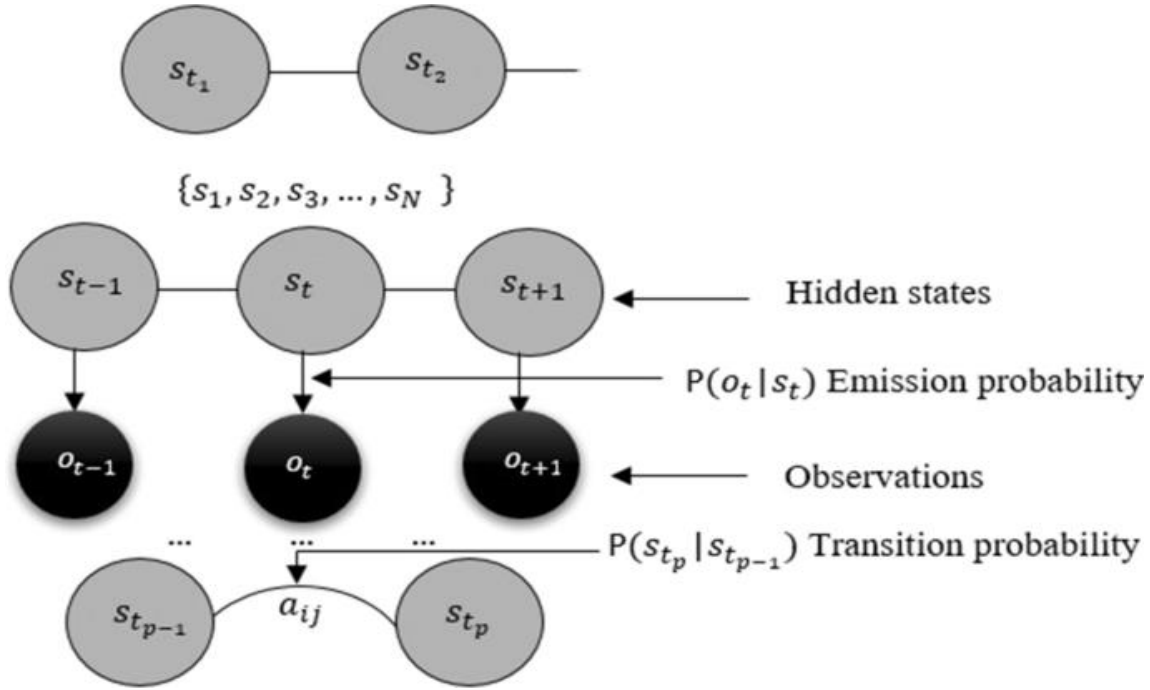

**Fig 1: Illustrates the graphical representation of a hidden Markov model, including hidden states, transition probabilities, observations, and emission probabilities (Sassi, 2021).**

**Parameters of Hidden Markov Model**

Hidden Markov Model (HHM) is basically structured with the following parameters.

1. The number of states in the model, sometimes referred to as the set of N hidden states ($S^N$), is the total number of states that the underlying hidden Markov process has: $S_1, S_2, \ldots, S_N$.
2. The number of unique observations, $M$. The quantity of discrete observation symbols is given by

$$V = V_1, V_2, \ldots, V_M \tag{1}$$

3. Transitional probability matrix (A): This $N \times N$ matrix's elements, where $i, j\ c\ 1, 2, \ldots, N$, describe the likelihood of transitioning from state $Z_{t-1,j}$ to $Z_{t,i}$ in a single step. This could be expressed as

$$a_{ij} - P(Z_{t+1} = S_j\ c\ Z_t = S_i) \tag{2}$$

In a matrix, we frequently list the transition probabilities. The state transition matrix, also known as the transition probability matrix, is typically represented by $P$. The state transition matrix, if the states are $1, 2, \ldots, r$ is given by

$$P = \begin{bmatrix} P_{11} & P_{12} & \cdots & P_{1r} \\ P_{21} & P_{22} & \cdots & P_{2r} \\ \vdots & \vdots & \vdots & \vdots \\ P_{r1} & P_{r2} & \cdots & P_{rr} \end{bmatrix} \tag{3}$$

Note that $P_{ij} \geq 0$, and for all $i$, we have

$$\sum_{k=1}^{r} P_{ik} = \sum_{k=1}^{r} P(X_{m+1} = k | X_m = i) = 1 \tag{4}$$

4. The observation likelihood or emission probability matrix $(B)$ is a $M \times N$ matrix whose members $B_{nj}$ $n \, \varepsilon \, N$ describe the likelihood of making an observation $X_{t,n}$ given a state $Z_{t,j}$. This could be expressed as

$$b_j(k) = P(x_t = v_t | Z_t = S_t) \tag{5}$$

An emission probability matrix of size $(M + 2) \times (N + 2)$ is given as

$$P = \begin{bmatrix} b_0(k_0) & \cdots & \cdots & \cdots & \cdots & \cdots & \cdots & \cdots \\ \cdots & b_1(k_1) & b_2(k_1) & b_3(k_1) & \vdots & \vdots & \vdots & b_N(k_1) & \cdots \\ \cdots & b_1(k_2) & b_2(k_2) & b_3(k_2) & \vdots & \vdots & \vdots & b_N(k_2) & \cdots \\ \cdots & \vdots & \vdots & \vdots & \vdots & \vdots & \vdots & \vdots & \cdots \\ \cdots & \vdots & \vdots & \vdots & \vdots & \vdots & \vdots & \vdots & \cdots \\ \cdots & \vdots & \vdots & \vdots & \vdots & \vdots & \vdots & \vdots & \cdots \\ \cdots & b_1(k_M) & b_2(k_M) & b_3(k_M) & \vdots & \vdots & \vdots & b_N(k_M) & \cdots \\ \cdots & \cdots & \cdots & \cdots & \cdots & \cdots & \cdots & \cdots & b_f(k_f) \end{bmatrix} \tag{6}$$

$b_j(k_j)$ is the probability of emitting vocabulary item $k_j$ from state $s_i$:

$$b_j(k_j) = P(O_t = k_j | X_t = s_i) \tag{7}$$

5. Initial probability distribution $\prod$: This is a $N \times 1$ vector of probabilities

$$\prod_j = p(z_1 = S_j) \tag{8}$$

**Solution to the Likelihood problem**

In order to solve the Likelihood problem, we determine the probability of an observation sequence $X = x_1, x_2, x_3$ given a model, or $P(x|\lambda)$. We took into account the state sequence $Q = q_1, q_2, q_3$, where $q_1$ and $q_3$ represent the beginning and final states, respectively. For our state sequence $Q$ and a model $\lambda$, the probability of an observation $X$ series can be expressed as

$$P(X | Q, \lambda) = \prod_{t=1}^{N} P(x_t | q_t, \lambda) = b_{x1}(q_1) \cdot b_{x2}(q_2) \cdot b_{x3}(q_3) \cdot b_{x4}(q_4) \tag{9}$$

From the property of a Markov chain, we represented the probability of the state sequence as

$$P(Q|\lambda) = \pi_{q1} \cdot p_{q1A2} \cdot p_{q2A3} \cdot p_{q3A4} \tag{10}$$

Summing over all possible state sequences is as follows:

$$\mathbb{P}(X|\lambda) = \sum_Q \mathbb{P}(X, Q|\lambda) = \mathbb{P}(X|Q, \lambda) \cdot \mathbb{P}(Q|\lambda) \tag{11}$$

$$\mathbb{P}(X|\lambda) = \sum_Q \pi_{q1} \cdot b_{x1}(q_1) \cdot p_{q1.q2} \cdot b_{x2}(q_2) \cdot p_{q2.q3} \cdot b_{x3}(q_3) \cdot b_{x4}(q_4) \tag{12}$$

**Forward Algorithm**

Using a forward variable, $\alpha_t(i)$ which indicates the likelihood that, given the HMM model $\lambda$, a partial observation sequence will occur up to time $t_{(i)}$ and the underlying Markov process will be in the state $T_{(i)}$.

$$a_{t^{(i)}} = \mathbb{P}(x_1, x_2, x_3, x_4 = T_i|\lambda) \tag{13}$$

We compute $\alpha_t(i)$ recursively via the following steps:

1. Initialize the forward probability as a joint probability of state $T_i$ and initial observation $m_1$. Let $\alpha_1(i) = \pi_i b_i(\pi_i)$ for $1 \leq i \geq 4$.
2. Compute $\alpha_3(j)$ for all states $j$ at $t = 3$, using the induction procedure, substituting $t = 1, 2, 3$:

$$\alpha_{t+1}(j) = \left[\sum_{i=1}^N \alpha_t(i) \cdot p_{ij}\right] b_j(x_t), 1 \leq t \leq (n-1), 1 \leq j \leq N \tag{14}$$

3. Using the results from the preceding step, compute $\mathbb{P}(x|\lambda) = \sum_{j=1}^N a_4(j)$.

**Backward Algorithm**

This research also constructs a backward variable for the forward algorithm, $\beta_i(i)$, which indicates the likelihood of a partial observation sequence from time $t + 1$ to the end (instead of up to t as in the forward method), where the Markov process is in state $s_1$ at time $t$ for a specific model, $\lambda$. The backward variable can be modeled mathematically as

$$\beta_i(i) = \mathbb{P}(x_{t+1}, x_{t+2}, x_{t+3}, x_{t+4}|q_{t,} = s_{i,}\lambda). \tag{15}$$

You can compute $a_t(i)$ recursively via the following steps:

1. Define $\beta_n(i) = 1$ for $1 \leq i \geq N$.

2. Compute $\beta_t(i) = \sum_{j=1}^N p_{ij} b_j(x_{t+1})\beta_{t+1}(j)$.

**Solution to Decoding problem**

Viterbi algorithm says that to find the state that maximizes the conditional distribution of states given that data is provided a function μ it depends on the previous step, the transition probabilities and the emission probabilities.

The following equations represent the highest probability along a single path for first $t$ observations which ends at state $i$.

$$w_i(t) = \max s_1, \ldots, s_{t-1} p(s_1, s_2, s_3, s_4 = i, v_1, v_2, v_3, v_4 | \theta) \qquad (16)$$

Using the same approach as the forward algorithm, we can calculate $w(t+1)$

$$w_i(t+1) = max_i(w_i(t) a_{ij} b_{jkv}(t+1)) \qquad (17)$$

To find the sequence of the hidden states, we need to identify the state that maximizes $w(t)$ at each time step $t$.

$$\arg max_t w(t)$$

**Solution to the Learning problem**

**Baum-Welch Algorithm**

The Baum-Welch Algorithm, also known as the Forward-Backward Algorithm or the Expectation-Maximization (EM) Algorithm for Hidden Markov Models (HMMs), is used to estimate the unknown parameters of an HMM. The algorithm has two steps, the E-step and the M-step. These steps are:
1. Initialize A and B
2. Iterate until convergence.

**E-Step**
- $\xi_i(t) = \dfrac{a_i(t) a_{ij} b_{jkv}(t+1) \beta(t+1)}{\sum_{i=1}^{M} \sum_{j=1}^{M} a_i(t) a_{ij} b_{jkv}(t+1) \beta(t+1)}$ (18)
- $\gamma_i(t) = \sum_{j=1}^{M} \xi_i(t)$ (19)

**M-Step**
- $\widehat{a_{ij}} = \dfrac{\sum_{t=1}^{T} \xi_i(t)}{\sum_{t=1}^{T} \sum_{j=1}^{M} \xi_i(t)}$ (20)
- $b_{jk} = \dfrac{\sum_{t=1}^{T} \widehat{\gamma_j(t)} 1(v(t)=k)}{\sum_{t=1}^{T} \gamma_j(t)}$ (21)
- Return A, B

## Long short-term memory (LSTM)

The Long Short-Term Memory (LSTM) is a Recurrent Neural Network (RNN) architecture designed to address the disappearing and exploding gradient problem. It uses memory cells, gates (input, output, and forget gates), and a hidden state to recognize long-term dependencies in sequential data. LSTMs are widely used in natural language processing, speech recognition, robotics, finance, and healthcare. A crucial component of LSTM is the "cell state," which preserves information over time. The gating mechanism, implemented using sigmoid functions, decides which data to ignore or retain based on the previous output and new input.

For each gate in the network, the learnable weights W (input weights), R (recurrent weights), and b (bias) are individually initialized as a column matrix, as given in Eq22.

$$W = \begin{bmatrix} W_i \\ W_f \\ W_g \\ W_o \end{bmatrix}, R = \begin{bmatrix} R_i \\ R_f \\ R_g \\ R_o \end{bmatrix}, b = \begin{bmatrix} b_i \\ b_f \\ b_g \\ b_o \end{bmatrix} \qquad (22)$$

The cell state and concealed state at time 't' are provided by Eq23. The Hadamard product is represented by $\odot$ and $\sigma_c$ is the state activation function-hyperbolic tangent function (tanh) respectively (element-wise multiplication).

$$C_t = f_t \odot C_{t-1} + i_t \odot g_t \qquad (23)$$

$$H_t = o_t \odot \sigma_c(C_t) \qquad (24)$$

At each timestep 't', equation 23 illustrates the update process of values in the network. The activation function for the gates, denoted as 'g', is a sigmoid function.

$$i_t = \sigma_g(W_i * x_i + R_i H_{t-1} + b_i) \qquad (25)$$

$$f_t = \sigma_g(W_f * x_i + R_f H_{t-1} + b_f) \qquad (26)$$

$$g_t = \sigma_c(W_g * x_i + R_g H_{t-1} + b_g) \qquad (27)$$

$$o_t = \sigma_g(W_o * x_i + R_o H_{t-1} + b_o) \qquad (28)$$

Gates in LSTMs control the addition and removal of information from the cell state. These gates may allow information to enter and exit the cell. It has a sigmoid neural network layer and a pointwise multiplication operation that help the mechanism (Vishal Vijayshankar Mishra, 2017).

**Methods**

This study information was obtained from the Nigeria index Mundi website and the central bank of Nigeria statistics database website on climate change and the production of maize. The factors taken into account were the annual temperatures, annual rainfall, and the annual production of maize in Nigeria from 1990 to 2022.

In this experiment, the performance of the models will be evaluated using some measures. The Hidden Markov Model (HMM) was then trained using the Forward-Backward Algorithm and then predicted the most likely states with Viterbi Algorithm and also forecasted using Baum Welch Algorithm. The Long short-term memory model (LSTM) was trained also using the means square error, the ADAM optimizer was used for the weight updates, and 200 epochs, where one epoch looped over all batches once. The best model due to their performance comparison was used to forecast.

**Experimental results and discussion**

In analyzing the result, we contrast the effectiveness of the Long short-term model (LSTM) network and the Hidden Markov Model (HMM) technique. The datasets was partitioned so that we could train and test our models using 80% and 20% of the same data set.

Measures for the predictions of both models using various metrics are shown in the results reported in the following subsection. The values of the measures for both models are shown in table 1.

Table 1: This table shows the values for all the measures for the prediction of both models.

|      | MAPE  | RMSE | Corr  | SEM  | MSE  |
|------|-------|------|-------|------|------|
| **HMM**  | 1.26  | 0.37 | -0.85 | 0.18 | 0.13 |
| **LSTM** | 12.98 | 1.21 | -0.85 | 4.19 | 0.87 |

HMM performed better than LSTM according to the study's conclusions. HMM had lower MAPE and RMSE values (1.26 and 0.37) compared to LSTM (12.98 and 1.21). HMM also had lower SEM and MSE values (0.18 and 0.87) compared to LSTM (4.19 and 0.87). The two models showed a negative correlation of -0.85.

**LSTM training and evaluation analysis on Maize production**

The input layer passes the data onto the LSTM layer that has 200 nodes at the output. The output of the LSTM layer is passed onto a dense layer 1 (one) that has 200 nodes at its input. The dense layer 1 uses mean square error (MSE) as the loss function and ADAM as the optimizer. The dense layer 1 is finally connected to the output layer that is also a fully-connected layer.

Table 2: This table shows some parameters of the LSTM in training and evaluation

| HIDDEN LAYER | NODE or NEURONS | EPOCHS | LOSS | VAL_LOSS | TEST ACCURACY |
|---|---|---|---|---|---|
| 1 | 32 | 200 | 0.7100 | 0.1055 | 0.5813 |

From table 2, the training loss is 0.7100, indicating reasonable performance on the training data. The validation loss is 0.1055, lower than the training loss, indicating good generalization to new data. The test accuracy is 0.5813, suggesting poor performance on test data and limited generalization. In comparing the HMM with LSTM for predicting maize yield, the HMM outperforms LSTM. Therefore, the HMM model is used for future predictions.

**Application of Hidden Markov Model for Maize Yields Forecast**

Hidden Markov Models are thought of as an unsupervised learning process where just the observed symbols are visible and the number of hidden states is unknown. In order to anticipate future years, this section uses Nigeria's maize yield statistics and a Hidden Markov model. The summary of the data is presented in Table3.

Table 3: The States and Observations of the Hidden Markov Model for a period of thirty-two years

| Year | States | Observation |
|---|---|---|
| 1990 | 1 | M |
| 1991 | 3 | M |
| 1992 | 1 | M |
| 1993 | 1 | M |
| 1994 | 3 | M |
| 1995 | 1 | M |
| 1996 | 3 | M |
| 1997 | 3 | L |
| 1998 | 2 | L |
| 1999 | 3 | M |
| 2000 | 1 | L |
| 2001 | 1 | M |
| 2002 | 1 | M |
| 2003 | 3 | M |
| 2004 | 1 | M |
| 2005 | 2 | M |
| 2006 | 2 | M |
| 2007 | 2 | M |
| 2008 | 3 | M |
| 2009 | 2 | M |

| 2010 | 4 | M |
|------|---|---|
| 2011 | 2 | M |
| 2012 | 3 | M |
| 2013 | 2 | M |
| 2014 | 2 | H |
| 2015 | 1 | H |
| 2016 | 4 | H |
| 2017 | 4 | H |
| 2018 | 4 | H |
| 2019 | 1 | H |
| 2020 | 4 | H |
| 2021 | 4 | H |

**Validity Test for the Model**

In analyzing the result of the Hidden Markov Model (HMM) process, there is a need to determine the transition matrix. Thus, the data was entered into R and the following results were produced for the transition matrix for both temperature and rainfall.

$$A = \begin{array}{c} \\ LL \\ LH \\ HL \\ HH \end{array} \overbrace{\begin{bmatrix} 0.3000 & 0.1000 & 0.4000 & 0.2000 \\ 0.1250 & 0.3750 & 0.3750 & 0.1250 \\ 0.5000 & 0.3750 & 0.1250 & 0.0000 \\ 0.2000 & 0.2000 & 0.0000 & 0.6000 \end{bmatrix}}^{LL \quad LH \quad HL \quad HH} \tag{28}$$

While Observation count matrix and Observation probability matrix or the emission probability matrix are given in equations (29) and (30), respectively.

$$D = \begin{array}{c} \\ LL \\ LH \\ HL \\ HH \end{array} \overbrace{\begin{bmatrix} 2 & 1 & 7 \\ 1 & 1 & 6 \\ 0 & 1 & 7 \\ 5 & 0 & 1 \end{bmatrix}}^{L \quad M \quad H} \tag{29}$$

$$B = \begin{array}{c} \\ LL \\ LH \\ HL \\ HH \end{array} \overbrace{\begin{bmatrix} 0.2000 & 0.1000 & 0.7000 \\ 0.1250 & 0.1250 & 0.7500 \\ 0.0000 & 0.1250 & 0.8750 \\ 0.8333 & 0.0000 & 0.1667 \end{bmatrix}}^{L \quad M \quad H} \quad (30)$$

The initial state probability is given below

$$\pi = \overbrace{0.3125 \quad 0.2500 \quad 0.25000 \quad 0.1875}^{LL \quad LH \quad HL \quad HH} \quad (31)$$

After 1000 iteration of Baum-Welch Algorithm, equation (32) settled to (33)

$$\lambda_1^* = (\hat{A}, \hat{B}, \hat{\pi}) \quad (32)$$

Where

$$\pi^* = \overbrace{0.2500 \quad 0.2500 \quad 0.2500 \quad 0.2500}^{LL \quad LH \quad HL \quad HH} \quad (33)$$

This $\pi^*$ shows the initialization process. The probabilities assigned is 0.2500 across the four (4) states.

$$\hat{A} = \begin{array}{c} LL \\ LH \\ HL \\ HH \end{array} \overbrace{\begin{bmatrix} 1.0000 & 0.0000 & 0.0000 & 0.0000 \\ 0.0924 & 0.8076 & 0.0000 & 0.1000 \\ 0.0000 & 0.2246 & 0.7754 & 0.000 \\ 0.0000 & 0.0000 & 0.2052 & 0.7948 \end{bmatrix}}^{LL \quad LH \quad HL \quad HH} \quad (34)$$

$$\hat{B} = \begin{array}{c} LL \\ LH \\ HL \\ HH \end{array} \overbrace{\begin{bmatrix} 0.0000 & 0.0000 & 1.0000 \\ 0.0000 & 1.0000 & 0.7809 \\ 0.0000 & 1.0000 & 0.0000 \\ 0.5272 & 0.4728 & 0.000 \end{bmatrix}}^{L \quad M \quad H} \quad (35)$$

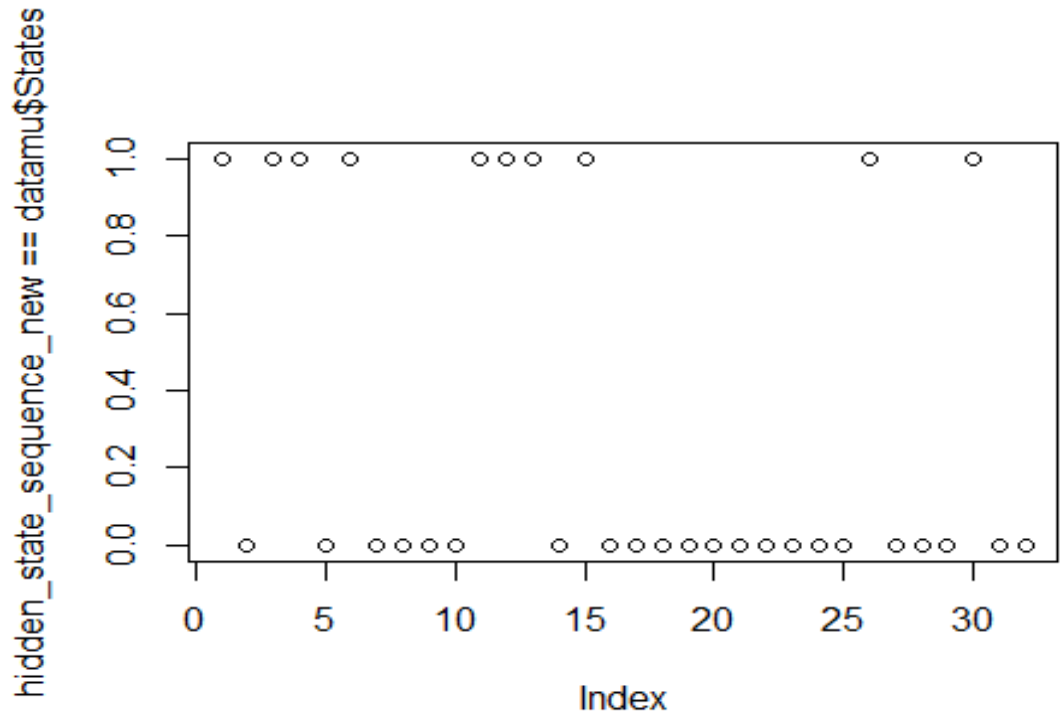

Fig 2: Plot comparing the model to the actual data

**Most Likely States Using Viterbi Algorithm**

In finding the most likely states based on A and B estimates from the observed data, the following results were gotten employing the Viterbi Algorithm.

3 1 3 1 3 1 1 3 1 1 3 1 3 1 3 1 3 1

Where state 1 is **Low Rainfall and Low Temperature** and state 3 is **High Rainfall and Low Temperature.**

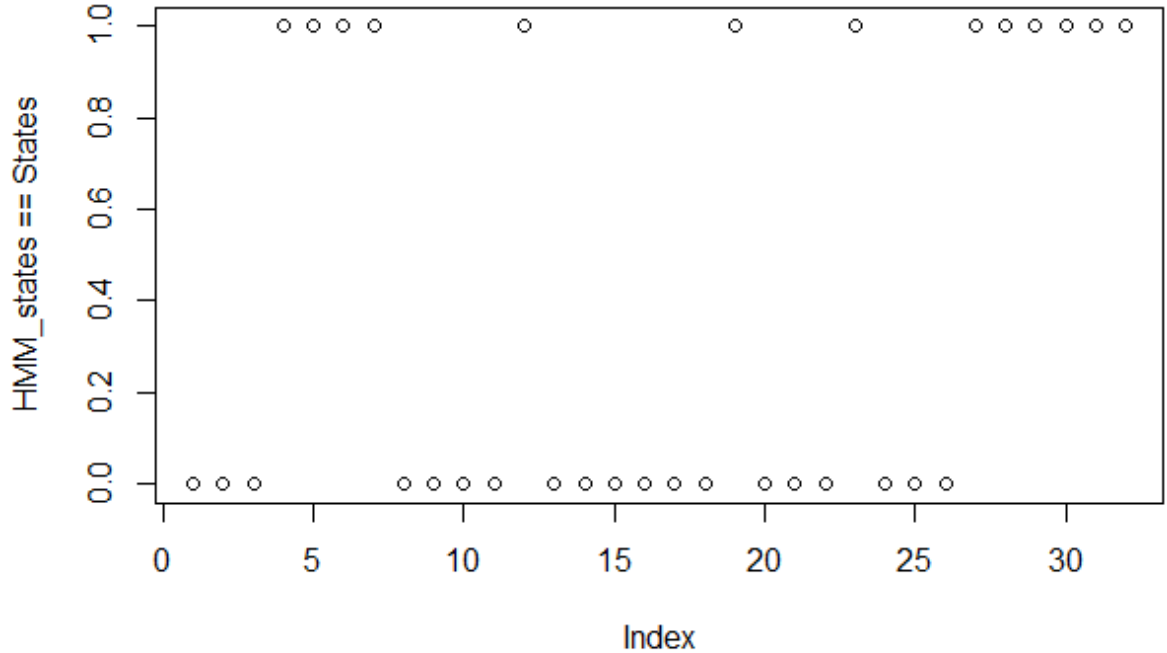

Fig3: This plot shows that the hidden states are solely responsible for the observable factors.

sum(hidden_state_sequence_new == datamu$States)/length(datamu$States)

0.3125

This shows that the hidden state (LL, LH, HL, HH) are 31.3% responsible for the observable factors of maize production (Low, Moderate, High).

**Steady State**

The likelihood that a Markov chain will remain in each state over the long run is its steady-state behavior. The initial probabilities are assumed to be equal for all states. The resulting steady state probabilities are:

$$\pi = \begin{matrix} LL & LH & HL & HH \\ [0.7738 & 0.1310 & 0.0952 & 0.0000] \end{matrix} \qquad (36)$$

Which indicate that in the long run the system will be in state 1 about 77% of the time, in state 2 about 13% of the time in state 3 about 9% of the time, and in state 4 never. This is consistent with the prediction made below.

**Hidden Markov Model for future forecast**

HMM was developed to forecast maize yield for future years, the parameters of the HMM were determined utilizing rainfall, temperature and maize yield data from 1990 to 2021, at the point, we made forecast for 2022, 2023, 2024, and 2025.

Future Forecast States and Observations

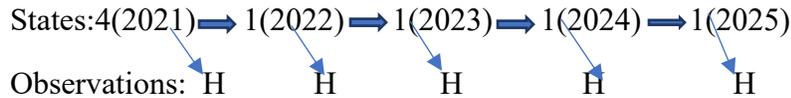

States: 4(2021) ⟹ 1(2022) ⟹ 1(2023) ⟹ 1(2024) ⟹ 1(2025)

Observations:  H         H         H         H         H

Baum Welch algorithm, $\lambda_1$ , settled to another model, $\lambda_1^*$, this new model was then used to make a forecast for future years. From the forecast, the HMM was in state 4 at time T (2021) emitting High rice maize, at that point, it then makes move to state 2 at time T+1 (2022) emitting High maize yield. Similar interpretation is given to move to state 2 at time T+2 (2023), move to state 2 at time T+3 (2024), move to state 2 at time T+4 (2025) all emitting High rice maize.

**Conclusion**

This experiment gives a quick overview of the HMM and LSTM, highlighting the salient features of stochastic modeling and deep learning for crop production. We also show the individual effectiveness of each model and the similarities between the HMM and LSTM using some multiple measures respectively. We discover from the measurement that there are some broad similarities, particularly at lower dimensionalities. We found out that HMM was the best model after checking their performances. Hidden Markov Model was then applied for future forecast on observed maize production data and use to develop a model which depicts sequences of observable factors given some hidden state and to estimate the predicted maize production in Nigeria.

The stable state probabilities were estimated alongside the transition and emission probabilities. The hidden state showed to be 21.9% responsible for the variables which can be observed and the estimated model predicts a 31.3% chance that the observable factors are influenced by the hidden states.

Given the results, the HMM is the higher performing model when predicting production of maize when comparing it with the LSTM. More than twice as low as the MAPE and RMSE of the LSTM forecast were the MAPE and RMSE of the HMM prediction. We say that the Hidden Markov Model is suitable for modelling maize production as it models either low production, moderate production and high production in Nigeria. It reveals the probability of changing from one state to another and the probability of emitting observations when in specific states.